\documentclass[twocolumn,floatfix]{revtex4}
\usepackage{amsmath}
\usepackage{amssymb}
\usepackage{psfig}
\usepackage{bm}

\begin{document}
\title{Detected jump-error correcting quantum codes, quantum error designs and quantum computation}
\author{G. Alber$^1$, Th. Beth$^2$, Ch. Charnes$^3$, A. Delgado$^4$, M. Grassl$^{2,5}$,
M. Mussinger$^1$}
\affiliation{$^1$ Institut f\"ur Angewandte Physik, Technische Universit\"at Darmstadt, D--64289 Darmstadt, Germany\\
         $^2$ Institut f\"ur Algorithmen und Kognitive Systeme, Universit\"at Karlsruhe, D--76128, Germany\\
         $^{3}$ Department of Computer Science \& Software Engineering, University of Melbourne, Parkville, Vic 3052, Australia\\
         $^4$ Department of Physics and Astronomy, University of New
         Mexico, Albuquerque, New Mexico 87131, USA\\
         $^5$ The Mathematical Sciences Research Institute,
         1000 Centennial Drive, Berkeley, CA 94720-5070, USA}
\date{August 14, 2002}

\begin{abstract}
The recently introduced detected-jump correcting quantum codes are
capable of stabilizing qubit-systems against spontaneous decay
processes arising from couplings to statistically independent
reservoirs.  These embedded quantum codes exploit classical
information about which qubit has emitted spontaneously and correspond
to an active error-correcting code embedded in a passive
error-correcting code.  The construction of a family of one
detected jump-error correcting quantum codes is shown and the optimal
redundancy, encoding and recovery as well as general properties of
detected jump-error correcting quantum codes are discussed.  By the use of
design theory multiple jump-error correcting quantum codes can be constructed.
The performance of one jump-error correcting quantum codes under non-ideal
conditions is studied numerically by simulating a quantum memory and
Grover's algorithm.
\end{abstract}

\maketitle

\flushbottom


\section{Introduction}

The discovery of powerful, fast quantum algorithms \cite{Shor1,Grover,Simon}
launched 
new efforts to implement such quantum algorithms in real physical systems.
Quantum algorithms
simultaneously exploit two characteristic features of quantum theory.
Namely, the fundamental phenomenon of quantum interference and
the fact that for distinguishable quantum systems the
dimension of the Hilbert space increases exponentially with the number of
systems. Therefore,  to
implement a quantum algorithm in a real quantum system
we must be able to create and manipulate arbitrary superpositions of quantum states
and to preserve quantum coherence during a computation. Unfortunately,
quantum coherence is very fragile. Typically, any coupling
to an environment 
leads to decoherence so that quantum mechanical superpositions are rapidly destroyed.

The urgent need to develop efficient methods to protect quantum coherence
has led to the study of very general classes of
quantum error-correcting codes \cite{Shor2,Knill1,Ekert,Knill2}.
The main idea is to restrict the dynamics of a quantum algorithm to
a subspace of the Hilbert space, in which errors can be
identified uniquely by suitable measurements and where the
error operators can be inverted by unitary operations.
Typically, this is achieved by an  encoding of the logical
information and by a suitable choice of quantum gates.

For some special cases it is also possible to design a passive
error-correcting quantum code \cite{Lidar1,Zanardi,Lidar2}.  Such a
passive quantum code relies on a subspace of the Hilbert space which
is not affected by any errors at all. In this situation the unitary
recovery operation is the identity operation so that an active
correction of the errors is not necessary.

In principle, any type of error can be corrected by these strategies
as long as enough physical qubits are available to achieve the
required redundancy and one can make a large number of control
measurements and perform the rapid recovery operations.  However, in
view of current-day experimental possibilities \cite{Sackett} it is
generally difficult to achieve both requirements.  Therefore it is
desirable to develop alternative error-correcting strategies which
possibly correct a restricted class of errors only, but which tend to
minimize both redundancy and the number of recovery operations.
Recently, the first steps in this direction have been taken by
defining a new class of one {\em detected jump-error correcting
quantum codes} which are capable of stabilizing distinguishable qubits
against spontaneous decay processes into statistically independent
reservoirs \cite{Alber01}.  These codes are constructed by embedding
an active error-correcting code in a passive code space and by
exploiting information available on error positions.  This embedding
procedure leads to a significant reduction of redundancy and the
number of control measurements and recovery operations.

In this paper the physical principles underlying detected jump-error
correcting quantum codes are explored and generalized, motivated by
the practical need for quantum error-correcting codes which minimize
both redundancy and the number of recovery operations.  Based on these
physical principles an upper bound is established on the number of
logical states of a general embedded detected jump-error correcting
quantum code.  

From this bound it is apparent that the recently
discovered one detected jump-error correcting quantum codes have
minimal redundancy.  Based on this family of optimal one {\em
detected jump-error correcting quantum codes}, we establish links with
the general notions of combinatorial design theory \cite{Beth1}.  For
this purpose the new concept of a {\em spontaneous emission error
design} is introduced.  This is a powerful tool for constructing
multiple detected jump-error correcting quantum codes capable of stabilizing
distinguishable qubits against spontaneous decay processes.
As an example, we present a new embedded three detected jump-error
correcting quantum code.

This paper is organized as follows. In Sec.~\ref{MasterEquation} basic
physical aspects concerning the spontaneous emission of photons by
qubit-systems are summarized. In Sec.~\ref{CodeDesign} the physical
principles are explored which lead to the construction of one detected
jump-error correcting quantum codes.  The conditions for general
detected jump-error correcting quantum codes are given in
Sec.~\ref{GeneralDJ}.  The links with combinatorial design theory are
established in Sec.~\ref{Designs}. Finally, in Sec.~\ref{NonIdeal}
numerical examples are presented which exhibit basic stability
properties of the optimal one detected jump-error correcting quantum
codes.

\section{Spontaneous decay of distinguishable qubits and photon detection}
\label{MasterEquation}

In this section we summarize 
basic facts about the dynamical description of a quantum system
interacting with initially unoccupied modes of the electromagnetic field.
These considerations
are the starting point for the
development of optimal strategies of error correction, which we pursue
in the subsequent sections.

\begin{figure}[hbt]
\centerline{\psfig{figure=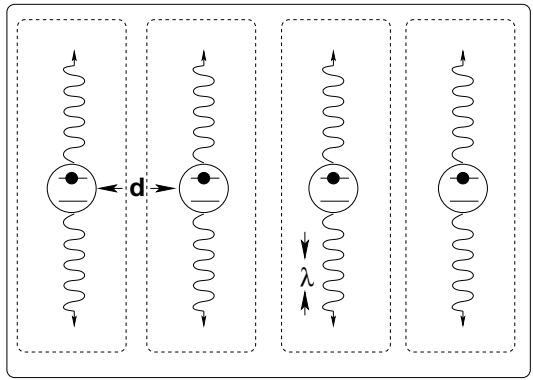,width=70mm}}
\caption{Model of a quantum computer consisting of an ensemble of
distinguishable two level systems (qubits) which are well localized at
positions ${\bf x}_{\alpha}$ and which are coupled to the vacuum modes of the
radiation field.  The mean distance ${\bf d}$ between adjacent qubits
is assumed to be large in comparison with the wave length of the
spontaneously emitted radiation, i.e.  $\lambda \ll {\bf d}$. Thus, the
spontaneous decay processes do not affect the distinguishability of
the qubits.  The qubits are monitored continuously by photodetectors
capable of determining the time at which a spontaneous decay process
occurred and the position of the affected qubit.\label{Iontrap2}}
\end{figure}

We consider a model of a quantum computer in which
$N$ two-level atoms (qubits) interact with external laser pulses
which synthesize the quantum gates underlying a quantum algorithm.
These $N$ qubits are assumed to be arranged in an array with
well defined  positions ${\bf x}_{\alpha}$ $(\alpha=1,\ldots,N)$ 
(see Fig.~\ref{Iontrap2}). In addition,
these $N$ qubits are assumed to be distinguishable, which requires that their mean
nearest neighbor distance is large in comparison with the optical
wave lengths involved. Their distinguishability guarantees that the dimension
of their associated Hilbert space is $K_{{\cal H}} = 2^N$ and thus scales
exponentially with the number of qubits.
In addition, it is assumed that these $N$ qubits
couple to the unoccupied modes of the electromagnetic field.
This coupling causes spontaneous decay processes of the qubits from
their excited states $|1\rangle_{\alpha}$, to their stable lower lying states
$|0\rangle_{\alpha}$.
Within the Born, Markov, and the rotating wave approximations the
resulting dynamics of the reduced density operator $\rho(t)$ of this
$N$-qubit system are described by the master equation
\cite{Mollow,Carmichael}
\begin{equation}
\frac{d\rho(t)}{dt}=-\frac{i}{\hbar}\left(H_{\rm eff}\rho(t)-\rho(t) H_{\rm eff}^{\dagger}\right)+
\sum_{\alpha=1}^{N}L_{\alpha}\rho(t) L_{\alpha}^{\dagger}
\label{master equation}
\end{equation}
with
the non-hermitian effective Hamiltonian
\begin{equation}
H_{\rm eff}=H-\frac{i \hbar}{2}\sum_{\alpha=1}^{N}L_{\alpha}^{\dagger}L_{\alpha}.
\label{effective Hamiltonian}
\end{equation}
Thereby, the coherent dynamics of the $N$-qubit system in the absence of the coupling to the
vacuum modes of the electromagnetic field are described  by the Hamiltonian $H$ which
incorporates the influence of the external laser pulses.
In addition, we assume that the
mean distance between the qubits is much larger than the wave lengths of the
spontaneously emitted radiation. Therefore,
to a good  degree of approximation each qubit $\alpha$ couples
to a different set of modes of the radiation field so that these sets constitute
statistically independent reservoirs \cite{Carmichael}. In Eq.~(\ref{master equation})
the coupling of qubit $\alpha$ to its reservoir and the resulting spontaneous decay process
$|1\rangle_{\alpha} \to |0\rangle_{\alpha}$ is characterized by the Lindblad operator
\begin{equation}
L_{\alpha} = \sqrt{\kappa_{\alpha}}~
\openone_{\beta\neq \alpha}
 \otimes |0\rangle_{\alpha \alpha} \langle 1 |
\label{jump operator}
\end{equation}
where $\openone_{\beta\neq \alpha}$ denotes the
identity on every except the $\alpha$-th qubit,
and $\kappa_{\alpha}$ is the associated spontaneous decay
rate.

Provided that initially the $N$-qubit system is in a pure state, say  $|\psi_0\rangle$,
a formal solution of the master Eq.~(\ref{master equation}) is given in \cite{Mollow}
\begin{eqnarray}
\rho(t) &\!\!=\!\!&\sum_{n=0}^{\infty}\int_0^t\!\!dt_n\int_0^{t_n}\!\!dt_{n-1}\ldots\int_0^{t_2}\!dt_1
p(t;t_{n}\alpha_{n},\ldots,t_1 \alpha_1)
\nonumber\\
&&
|t;t_{n}\alpha_{n},\ldots,t_1 \alpha_1\rangle \langle t;t_{n}\alpha_{n},\ldots,t_1 \alpha_1|
\label{unravel}
\end{eqnarray}
with the pure quantum state
\begin{eqnarray}
&&|t;t_{n}\alpha_{n},\ldots,t_1 \alpha_1\rangle\nonumber\\
&&\qquad=
\frac{
e^{-iH_{\rm eff}(t - t_n)/\hbar}
L_{\alpha_n}
\ldots
L_{\alpha_1}
e^{-iH_{\rm eff}t_1/\hbar}
}
{p(t|t_{n}\alpha_{n},\ldots,t_1 \alpha_1)^{1/2}
}|\psi_0\rangle\nonumber
\end{eqnarray}
and with the probabilities
\begin{eqnarray}
\label{probability}
&&p(t;t_{n}\alpha_{n},\ldots,t_1 \alpha_1) \\
&& =\langle \psi_0 |
e^{iH^{\dagger}_{\rm eff}t_1/\hbar}
L^{\dagger}_{\alpha_1}
\ldots
L^{\dagger}_{\alpha_n}
e^{iH^{\dagger}_{\rm eff}(t - t_n)/\hbar}\cdot\nonumber\\
&&\quad\qquad\qquad
e^{-iH_{\rm eff}(t - t_n)/\hbar}
L_{\alpha_n}
\ldots
L_{\alpha_1}
e^{-iH_{\rm eff}t_1/\hbar}|\psi_0\rangle .\nonumber
\end{eqnarray}
It can be shown that each pure state $|t;t_{n}\alpha_{n},\ldots,t_1
\alpha_1\rangle$ describes the quantum state of the $N$-qubit system
at time $t$ conditioned on the emission of precisely $n$ photons at
times $t_1 \le \ldots\le t_n$ by qubits $\alpha_1,\ldots,\alpha_n$
\cite{Mollow,Carmichael,Carmichael2,Zoller}. Thus, each of the pure
quantum states of Eq.~(\ref{unravel}) corresponds to a possible
measurement record $(t_1 \alpha_1,\ldots,t_n \alpha_n)$ in an
experiment in which each qubit is observed continuously by
photodetectors. In the subsequent discussion it is important to note
that due to the large separation between the qubits ideally this
measurement record not only determines the spontaneous decay times
$t_{i}$, but also the associated positions $\alpha_i$ ($i=1,\ldots,n$)
of the qubits which have been affected by these decay processes.  The
measurement record $(t_1 \alpha_1,\ldots,t_n \alpha_n)$ is observed
with probability $p(t;t_{n}\alpha_{n},\ldots,t_1 \alpha_1)$.
According to Eq.~(\ref{unravel}) the quantum state
$|t;t_{n}\alpha_{n},\ldots,t_1 \alpha_1\rangle$ resulting from a
particular measurement record is determined by two types of effects.
First, the time evolution between two successive photon emission
events is characterized by the non-hermitian Hamiltonian $H_{\rm eff}$
of Eq.~(\ref{effective Hamiltonian}).  Thus, even in the absence of
any spontaneous photon emission process in a given time interval
$[0,t]$, in general an arbitrary quantum state is modified by the
couplings to the vacuum modes of the electromagnetic field.  Second,
immediately after the spontaneous emission of a photon by qubit
$\alpha$ the quantum state is modified by the Lindblad operator
$L_{\alpha}$.  This distinction between the two different types of
dynamics is crucial for the development of a useful quantum
error-correcting strategy which minimizes the redundancy as well as
the number of control measurements and recovery operations.

\section{Designing optimal one detected jump-error correcting quantum codes} \label{CodeDesign} 
In this section we introduce the main ideas leading to the
construction of optimal one detected jump-error correcting embedded
quantum codes which are capable of stabilizing systems of
distinguishable qubits against spontaneous decay processes.  These
quantum codes exploit the physical possibilities which are offered by
the continuous observation of the distinguishable qubits by
photodetectors.

How can we stabilize an $N$-qubit system, such as the one depicted in
Fig.~\ref{Iontrap2}, against spontaneous decay processes if we can
observe the qubits continuously with the help of
photodetectors?  To achieve this two major
tasks have to be tackled. 
Namely, we have to correct the modifications taking place
during successive photon emission events.  These modifications are
described by the effective Hamiltonian of Eq.~(\ref{effective
Hamiltonian}).  And, we have to invert each quantum jump
$|1\rangle_{\alpha}\to |0\rangle_{\alpha}$ caused by the spontaneous
emission of a photon by qubit $\alpha$.

In principle, the errors 
taking place during two successive photon emission events can be corrected
by an active error-correcting quantum code with a sufficiently rapid sequence of control
measurements and recovery operations \cite{Pellizzari}. 
However
for practical applications such an approach is not very attractive. One faces
all the problems which are related to the implementation of
large numbers of control measurements and recovery operations.
It is therefore much more desirable to
develop passive error-correcting
methods for stabilizing the $N$-qubit system
against modifications arising 
from the effective Hamiltonian of Eq.~(\ref{effective Hamiltonian}).

The main idea is to determine an appropriate subspace $D$
of the Hilbert space of the $N$ distinguishable
qubits in which the perturbing part of the Hamiltonian acts as a multiple of the unit operator, i.e.
\begin{equation}
H_{\rm eff}|_{D} = H|_{D} - \frac{i}{2}c \openone |_{D}
\label{Heffective}
\end{equation}
where $c$ denotes a c-number. (In our setting 
this number is real-valued and positive.)  If the (unperturbed) dynamics
characterized by the Hamiltonian $H$ do not take an initial pure
quantum state outside this decoherence free subspace $D$,
 this condition implies that, provided no
photon is emitted in the time interval $[0,t]$, the quantum state at
time $t$ is identical with the unperturbed state (compare with
Eq.~(\ref{unravel})).  Thus, if one can find such a sufficiently high
dimensional decoherence free subspace, the dynamics taking place
between successive spontaneous photon emission events are stabilized
perfectly without the need for control measurements and recovery
operations. In practice it is desirable to choose the dimension of
the decoherence free subspace to be as large as possible. 

An important special case occurs when all the $N$ qubits have identical
spontaneous decay rates, i.e. $\kappa_{\alpha}\equiv \kappa$.  In this
situation it follows that
\begin{equation}\label{eq:constweight}
\sum_{\alpha =1}^{N}L_{\alpha}^{\dagger}L_{\alpha} \equiv
\kappa \sum_{\alpha =1}^{N}|1\rangle_{\alpha \alpha}\langle 1|
\end{equation}
and any subspace formed by basis states involving an equal number, say
$w\leq N$, of excited qubits is a decoherence free subspace.  For a
given number of $N$ qubits the dimension $K_{D}$ of such a
decoherence free subspace is given by $K_{D} = {N \choose w}$ which
is maximal if $w = \lfloor N/2 \rfloor$.  ($\lfloor x\rfloor$ denotes
the largest integer smaller or equal to $x$.)

In general, the first spontaneous emission of a photon 
will affect the quantum state of the $N$-qubit system in an irreversible way.
According to Eq.~(\ref{unravel}) the spontaneous emission of a photon  by qubit $\alpha$, for example,
is described by the application of the Lindblad operator $L_{\alpha}$ which induces
a quantum jump $|1\rangle_{\alpha}\to |0\rangle_{\alpha}$.
This Lindblad operator is not invertible over the decoherence free subspace $D$
so that this quantum jump cannot be corrected.
In order to correct for this quantum jump actively we have
to restrict the dynamics to a still smaller subspace $C \subset D$
in which a unitary operator, say $U_{\alpha}$, can be found 
having the property
\begin{equation}
U_{\alpha} L_{\alpha}|_{C} = \openone |_{C}.
\label{recovery}
\end{equation} 
Therefore, if we still want to take advantage of passive error
correction between successive photon emission events we have to
construct an active error-correcting quantum code within the relevant
decoherence free subspace.

We now construct a one detected jump-error correcting embedded
quantum code in the special case of identical spontaneous decay rates
considered above.  According to the criterion given in \cite{Knill1},
the orthogonal basis states $\{|{ \bm c}_i\rangle: i = 1,\ldots,K \}$
of a subspace $ C $ constitute an active error-correcting quantum code
with respect to the set of error operators $\{L_{\alpha}: \alpha =
1,\ldots,N \}$ if and only if
\begin{eqnarray}
 \langle {\bm c}_i| L^{\dagger}_{\alpha} L_{\beta} | {\bm c}_j\rangle
 = \delta_{ij}\Lambda_{\alpha \beta}
\label{Knill}
\end{eqnarray}
for all possible values of $i,j$ and $\alpha,\beta$.
Eq.~(\ref{Knill}) states the necessary and sufficient conditions for
the existence of unitary recovery operations which fulfill
Eq.~(\ref{recovery}) for the error operators $\{L_{\alpha}: \alpha =
1,\ldots,N \}$.  In the physical setting this criterion states that:
(i) different orthogonal quantum states remain orthogonal under the
action of error operators; and (ii) all basis states are affected by a
given pair of errors $L_{\alpha}$ and $L_{\beta}$ in a similar way.
The latter condition necessarily implies that the scalar products
between the states $L_{\alpha}|{\bm c}_i\rangle$ and $L_{\beta}|{\bm
c}_i\rangle$ are state-independent.  It is plausible that a larger set
of error operators leads to a more restrictive set of conditions of
the type of Eq.~(\ref{Knill}).  Furthermore, we also expect that more
restrictive conditions lead to a higher redundancy of an active
quantum code.

As an example, consider the  situation where continuous observation of the
$N$-qubit system by photodetectors does not reveal which qubit
$\alpha$ has emitted the registered photon.  This implies that the
error operators which could induce a spontaneous decay process are in
the set $\{L_{\alpha}: \alpha = 1,\ldots,N \}$.  It has been shown by
Plenio et al. \cite{Plenio} that when the error
positions are unknown, eight physical qubits are needed to encode two
orthogonal logical states by an embedded quantum code.  This should be
compared with the optimal active one-error correcting code using five
qubits \cite{5bit}.

Thus, the advantage offered by using an embedded quantum code, capable
of passively correcting errors between successive photon emission events
leads to a significant increase of redundancy in comparison
to purely active methods.  However, this disadvantage can be overcome
if besides knowing the error time, information about the error
position $\alpha$ is also available.  In principle, this information
can be obtained from continuous observation of the $N$-qubit system by
photodetectors as long as the mean distance between adjacent qubits is
large in comparison with the wave length of the spontaneously emitted
radiation.  For this purpose it is important that each
photon which is emitted by one of the qubits can be detected.

How can we construct a one detected jump-error correcting embedded quantum code
which exploits information about the error position in an optimal way
so that its redundancy is minimized?  Let us
concentrate again on our previously introduced example of identical
spontaneous decay rates. In this setting we have a decoherence free
subspace which involves $w$ excited qubits. This stabilizes the
dynamics between successive photon emission events passively.  For
example, in the simple case of $N=4$ and $w=2$, the orthogonal basis
states $|{\bm c}_i\rangle$ of the decoherence free subspace are given by
$\{|1100\rangle, |0011\rangle, |1010\rangle, |0101\rangle,
|1001\rangle,|0110\rangle\}$.  Within this six-dimensional decoherence
free subspace a possible active quantum code, capable of correcting
one quantum jump at a time at a known position $\alpha$, is determined
by the following three orthonormal code words:
\begin{eqnarray}
|{\bm c}_0\rangle &=& \frac{1}{\sqrt{2}}[|1100\rangle + e^{i\varphi}|0011\rangle],\nonumber\\
|{\bm c}_1\rangle &=& \frac{1}{\sqrt{2}}[|01010\rangle + e^{i\varphi}|1010\rangle],\nonumber\\
|{\bm c}_2\rangle &=& \frac{1}{\sqrt{2}}[|1001\rangle + e^{i\varphi} |0110\rangle].
\label{jump1}
\end{eqnarray}
These code words are formed by all possible different pairs of basis
states of the decoherence free subspace which can be constructed by
interchanging states $|0\rangle$ and $|1\rangle$.  These complementary
pairings are also characterized by equally weighted probability
amplitudes which involve an arbitrary phase $\varphi$.  It can be
easily checked that this embedded quantum code fulfills the criterion
of Eq.~(\ref{Knill}) for any of the Lindblad operators $L_{\alpha}$ of
Eq.~(\ref{jump operator}) separately, but not for two different
Lindblad operators with $\alpha \neq \beta$.  Thus, provided that the
error time and the error position are known this embedded quantum code can
correct one quantum jump at a time.  In addition, errors between
successive quantum jumps are corrected passively.  
In this way the
number of control measurements and recovery operations is reduced and
so is the redundancy. This is apparent by comparing our quantum code
with the code proposed by Plenio et al. \cite{Plenio}, which requires
eight physical qubits for the encoding of two orthogonal quantum
states.

The above construction based on complementary pairings can be
generalized to an even number $N$ of physical qubits;  giving an
infinite family of one detected jump-error correcting embedded quantum codes.
The construction is the following.  We start from the highest
dimensional decoherence free subspace which involves all quantum
states in which half of the qubits are excited to states
$|1\rangle_{\alpha}$.  Subsequently, we construct an active quantum
code within this decoherence free subspace by complementary pairings.
The dimension $K_{\rm J}$ of the resulting embedded quantum code is
given by
\begin{eqnarray}
K_{{\rm J}} &=& (1/2){N\choose N/2} \equiv {N-1 \choose N/2 -1}.
\end{eqnarray}
It is a remarkable fact that, provided one wants to correct errors passively
between successive quantum jumps, 
it is not possible for an even number of
physical qubits $N$ to reduce the redundancy of such
an embedded jump code (see Sec.~\ref{Designs} and Ref. \cite{DCC}). 
Therefore, our family of embedded quantum codes has the desirable feature that it
minimizes redundancy and that it requires only a small number of
recovery operations. For instance, 
in a given time interval, say $[0,t]$, the mean
number $n$ of required recovery operations is of the order of $ n
\approx 2t/(N \kappa)$.

To stabilize any quantum algorithm against spontaneous decay processes
using an embedded one detected jump-error correcting quantum  
code three requirements have to be met.  First, one has to be able to
register the time $t$ and position $\alpha$ of each spontaneous decay
event which takes place during the performance of the quantum
algorithm. As indicated schematically in Fig.~\ref{Iontrap2} this can
be achieved by continuous observation of the $N$-qubit system with 
photodetectors. 
In principle, an identification of the
perturbed qubit $\alpha$ is possible provided the mean nearest
neighbor spacing of the qubits is large in comparison with the wave
lengths of the radiation emitted spontaneously.  However, in practice
the error position $\alpha$ might not be determined so easily due to
imperfect detection efficiencies of the photodetectors.  Therefore, in
actual applications shelving techniques \cite{Dehmelt,Toschek} might
be useful which amplify each spontaneously emitted photon to such an
extent that it can be detected with an efficiency arbitrarily close to
unity. 

Second, we have to ensure that each spontaneous decay event is
corrected immediately by application of an appropriated unitary
transformation which inverts the effect of the Lindblad operator
$L_{\alpha}$.  In practice, this inversion has to be performed on a
time scale which is small in comparison with the natural time scale of
the quantum algorithm and with the mean spontaneous decay time.
Third, one has to ensure that the sequence of quantum gates which
constitute the quantum algorithm does not leave the code space at any
time. This can be done by encoding the logical information within the
code space, and to develop a universal set of quantum gates which
leaves this code space invariant.  

Ideally these quantum gates are
implemented by suitable Hamiltonians. This ensures that the code space
is left invariant even during the application of one of these
universal quantum gates.  Such universal sets of Hamiltonian-induced
quantum gates have already been developed for decoherence free
subspaces $D$ of the kind discussed above
\cite{Lidar1,Zanardi,Lidar2}.  But in general, unitary gates based on
swapping Hamiltonians need not be universal on the embedded quantum
code, or the swapping Hamiltonians do not leave the embedded quantum
code invariant. The solution of this intricate and yet unsolved
problem is beyond the scope of the present work.  However, some preliminary
results have already been obtained recently
~\cite{Fortschr,QIT,Khodjasteh}.

\begin{figure}[hbt]
\centerline{\psfig{figure=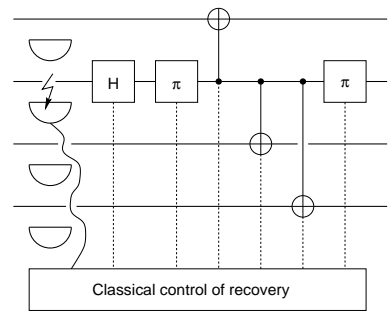,width=50mm}}
\caption{Gates to invert a quantum jump (indicated by a bolt) 
of a physical qubit. The four physical
qubits involved form a $(4,3,1)_2$ code. \label{gate}}
\end{figure}

So far we have shown that any Lindblad operator $L_{\alpha}$ of the
form of Eq.~(\ref{jump operator}) can be inverted by our one detected
jump-error correcting quantum codes.  We provide an example of a unitary
transformation which achieves this inversion in the case of the one
detected jump-error correcting quantum code involving four physical qubits.  A
possible sequence of quantum gates capable of inverting a
spontaneous decay process affecting qubit $\alpha =2$, for example, is
depicted in Fig.~\ref{gate}.  This example demonstrates the basic fact
that it is indeed possible to perform a unitary inversion of the
Lindblad operator $L_2$ provided Eq.~(\ref{Knill}) is fulfilled for
$\alpha = \beta =2$.

\section{General Detected Jump-Error Correcting Quantum Codes}
\label{GeneralDJ}
To define general detected jump-error correcting quantum
codes, we introduce some notation. For a set of positions
$E=\{\alpha_1,\alpha_2,\ldots,\alpha_n\}$, we denote by ${\cal J}_E$
the operator
$$
{\cal J}_{E}:=
{\cal J}_{\{\alpha_1,\alpha_2,\ldots,\alpha_n\}}:=
L_{\alpha_1}L_{\alpha_2}\ldots L_{\alpha_n}.
$$
The associated error times $t_1,\ldots,t_n$ are no longer mentioned
explicitly, but it is understood that they are known. Note that the
operators $L_{\alpha_1}, \ldots, L_{\alpha_n}$ commute, because the
$\alpha_i$ are pairwise different. Since 
by  Eq.~(\ref{probability}), the errors which involve two equal
indices, say $\alpha_1 = \alpha_2$, cannot occur. 

As discussed in Sec.~\ref{CodeDesign}, for $H\equiv 0$ all the states
which are superpositions of states with a constant number of excited
qubits are common eigenstates of the non-unitary effective time
evolution (\ref{Heffective}) between quantum jumps.  A
subspace $C$ of such a decoherence free subspace with orthonormal
basis $\{|{\bm c}_1\rangle,\ldots,|{\bm c}_K\rangle\}$ is called a
$d$-detected jump-error correcting quantum code, and is denoted by
$C=(N,K,d)_w$ if the following condition holds for sets $E$ of jump
positions with at most $d$ elements and for all basis states $|{\bm
c}_i\rangle$ and $|{\bm c}_j\rangle$:
\begin{equation}\label{eq:DJCcondition}
\langle{\bm c}_i|{\cal J}_E^\dagger{\cal J}_E|{\bm c}_j\rangle
 =\delta_{ij} \lambda(E).
\end{equation}
The notation $C= (N,K,d)_w$ is motivated by classical
coding theory. Similarly, the notation $d$-JC$(N,w,K)$
is motivated by notations from design theory
(see \cite{Alber01}).

The validity of  this statement follows from the general conditions on quantum
error-correcting codes (cf. \cite{Knill1}). Since we
know on which positions the jump operator ${\cal J}_E$ acts, only
products of the form ${\cal J}_{E}^{\dagger}{\cal J}_{E}$ have to be
considered.

There is a natural connection with combinatorics. For a basis state
$|x_1\ldots x_N\rangle$ of $N$ qubits, the positions which are in
state $|1\rangle$ define a subset of $\{1,\ldots,N\}$. Furthermore, a
collection of such subsets corresponds to an equally weighted
superposition of basis states.

Let ${\cal B}^{(1)},\ldots,{\cal B}^{(K)}$ be $K$ disjoint sets of subsets
$X\subseteq\{1,\ldots,N\}$, where each subset $X$ contains $w$
elements. Identifying the set $X$ and the binary word ${\bm x}$ where
${\bm x}_i=1$ if $i\in X$ and ${\bm x}_i=0$ otherwise, we define the
states
\begin{equation}\label{eq:combDJC}
|{\bm c}_i\rangle=\frac{1}{\sqrt{|{\cal B}^{(i)}|}}
                  \sum_{X\in{\cal B}^{(i)}}|{\bm x}\rangle.
\end{equation}
Where $|{\cal B}^{(i)}|$ denotes the number of elements
(cardinality) of the set ${\cal B}^{(i)}$.  These orthonormal states
span a $d$-detected jump-error correcting quantum code $C=(N,K,d)_w$
provided that for all sets of jump positions $E$ with no more than $d$
elements and all sets ${\cal B}^{(i)}$ the following condition holds:
\begin{equation}\label{eq:combDJCcondition}
\frac{|X\in {\cal B}^{(i)}: E\subseteq X\}|}
                       {|{\cal B}^{(i)}|}
                 =\lambda(E).
\end{equation}
We note that the disjointness of the sets ${\cal B}^{(i)}$ implies
condition (\ref{eq:DJCcondition}) for $i\ne j$.  Rewriting the
operator ${\cal J}_{E}^{\dagger}{\cal J}_{E}$ as $ {\cal
J}_{E}^{\dagger}{\cal J}_{E} =\sum_{X\supseteq
E}|{\bm x}\rangle\langle {\bm x}|$ shows that for $i=j$ and the states
(\ref{eq:combDJC}) the expectation value $\langle{\bm c}_i|{\cal
J}_E^\dagger{\cal J}_E|{\bm c}_i\rangle$ equals the expression in
(\ref{eq:combDJCcondition}).

\section{Block Designs and Jump Codes}\label{Designs}
In this section we will show that $d$-detected jump-error correcting
quantum codes are naturally connected with $d$-designs. These are
combinatorial structures which have been extensively studied for many
decades, cf.  \cite{Beth1}. To denote this class of combinatorial
structures we introduce some notations using the language of finite
incidence structures. Let $V$ be a set of $N$ elements, called {\em
points}, say $V=\{1,\ldots,N\}$ where $N>0$ is an integer and
$|V|=N$. For $0\le w\le N$ the class of $w$-subsets of $V$ containing
$w$ elements will be denoted by ${V \choose w}$. In a suggestive way
its cardinality, i.e.
$$
\left|{V \choose w}\right|={{|V|} \choose w}={N \choose w},
$$
is just the binomial coefficient $ {N \choose w}.  $ An incidence
structure in $V$ is specified by a distinguished class ${\cal B}$ of
subsets of $V$. The elements $B$ of ${\cal B}$ are called {\em blocks}
(or sometimes {\em lines}) of the incidence structure. If ${\cal
B}\subseteq {V \choose w}$, we say that ${\cal B}$ has constant block
size $w$. As an example, any undirected graph is an incidence
structure of block size two, if we choose $V$ as the set of points of
the graph and the points which are directly connected by an edge as
blocks.

\begin{figure}[hbt]
\centerline{\psfig{figure=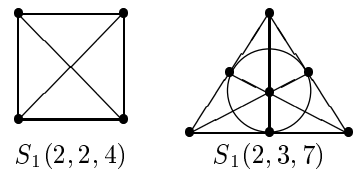}}
\caption{Examples of $2$-designs.\label{fig:designs}}
\end{figure}

For any point $\alpha\in V$ the class ${\cal B}_\alpha:=\{B\in{\cal
B}:\alpha\in B\}$ denotes the class of blocks containing the point
$\alpha$ (or: ``the lines through $\alpha$''). If $|{\cal
B}_\alpha|=r$ is constant for all $\alpha\in V$ the incidence
structure ${\cal B}$ is called regular. For a graph, $|{\cal
B}_\alpha|$ is the degree $d(\alpha)$ of the vertex $\alpha$, i.e.,
the number of edges on which $\alpha$ lies. The incidence structure
${\cal B}$ ---as well as the graph itself---is called regular, if
$d(\alpha)$ is constant. If there exists a constant $\lambda_d$ such
that for all $E=\{\alpha_1,\ldots, \alpha_d\}\in{V \choose d}$ the
class ${\cal B}_E:=((({\cal
B}_{\alpha_1})_{\alpha_2})\ldots)_{\alpha_d}$ of all blocks containing
the set $d$ of points has size $|{\cal B}_E|=\lambda_d$, the incidence
structure is called {\em $d$-regular}. A {\em $d$-design} is a
$d$-regular incidence structure with constant block size $w$. It is
denoted by $d$-$(N,w,\lambda_d)$-design or as $S_{\lambda_d}(d,w,N)$
\cite{Beth1}. Graphs which correspond to $2$-designs are depicted in
Fig.~\ref{fig:designs}.

The preceding discussion leads to the notion of $d$-spontaneous
emission error designs which we denote by $d$-${\rm SEED}(N,w,K)$
\cite{DCC}. The essential property of these combinatorial objects is
the {\em local multiplicity} $\lambda(E)$ of a subset $E$ of
$\{1,\ldots,N\}$ containing at most $d$ elements which is defined by
\begin{equation}\label{eq:localMult}
\frac{|{\cal B}^{(i)}_E|}{|{\cal B}^{(i)}|}=\lambda(E),
\end{equation}
where ${\cal B}^{(1)},\ldots,{\cal B}^{(K)}$ are disjoint subsets of
${V \choose w}$. Any $d$-${\rm SEED}(N,w,K)$ produces a
$d$-detected jump-error correcting quantum code $C=(N,K,d)_w$ using
the encoding defined in Eq.~(\ref{eq:combDJC}).

We conclude this section by 
constructing of a three-jump
correcting code. The permutation group
$G=\langle (1\,2)(3\,4), (1\,4)(2\,3),
(5\,6)(7\,8), (5\,8)(6\,7),(1\,2\,3)(5\,6\,7)\rangle$ 
of order $48$ acts on the $4$-element 
subsets of $\{1,2,\ldots,8\}$. 
The orbits under $G$ of the sets $\{1,2,5,6\}$, $\{1,3,5,6\}$, and
$\{1,4,5,6\}$ are mutually disjoint. Direct calculation shows that
they fulfill the local multiplicity condition
(\ref{eq:localMult}). Hence the sets 
${\cal B}^{(1)}:=\{1,2,5,6\}^G$,
${\cal B}^{(2)}:=\{1,3,5,6\}^G$, and 
${\cal B}^{(3)}:=\{1,4,5,6\}^G$ 
define an $(8,3,3)_4$ jump code. The corresponding (not normalized) basis states are
given by:

{\small
\begin{eqnarray*}
|c_1\rangle&\!\!=\!\!
&|00110011\rangle
+|00111100\rangle
+|01010101\rangle
+|01011010\rangle\\
&&\!{}+|01100110\rangle
+|01101001\rangle
+|10010110\rangle
+|10011001\rangle\\
&&\!{}+|10100101\rangle
+|10101010\rangle
+|11000011\rangle
+|11001100\rangle
\end{eqnarray*}
\begin{eqnarray*}
|c_2\rangle&\!\!=\!\!
&|00110110\rangle
+|00111001\rangle
+|01010011\rangle
+|01011100\rangle\\
&&\!{}+|01100101\rangle
+|01101010\rangle
+|10010101\rangle
+|10011010\rangle\\
&&\!{}+|10100011\rangle
+|10101100\rangle
+|11000110\rangle
+|11001001\rangle
\end{eqnarray*}
\begin{eqnarray*}
|c_3\rangle&\!\!=\!\!
&|00110101\rangle
+|00111010\rangle
+|01010110\rangle
+|01011001\rangle\\
&&\!{}+|01100011\rangle
+|01101100\rangle
+|10010011\rangle
+|10011100\rangle\\
&&\!{}+|10100110\rangle
+|10101001\rangle
+|11000101\rangle
+|11001010\rangle
\end{eqnarray*}
}

Further examples of $d$-${\rm SEED}$s are discussed in \cite{DCC}. In
that paper, there are also general bounds on the parameters of jump
codes derived. In particular, the dimension $K$ of a
$d$-detected jump-error correcting $(N,K,d)_w$ code is bounded above by
\begin{equation}\label{eq:gen_upperbound}
K \le \min\bigl\{{N-d \choose w-d},{N-d \choose w}\bigr\}.
\end{equation}
For completeness, we repeat the main ideas of the proof. 

The
dimension of the space spanned by the basis states of $N$ qubits where
$w$ of them are in the excited state is ${N \choose w}$. This implies
the bound $K\le{N \choose w}$. A jump on $j$ positions reduces the
number of excitations to $w-j$. After the jump, the $j$
positions where the jump occurred are zero. There are ${N-j \choose
w-j}$ such basis states. A jump must not reduce the dimension of the
code, hence $K\le {N-j \choose w-j}$. For $0\leq j \leq d$ possible
quantum jumps the lowest upper bound is achieved for $j=d$,
as
$$
{N \choose w}
=\frac{N(N-1)\ldots(N-d+1)}{w(w-1)\ldots(w-d+1)}{N-d \choose w-d}.
$$
To obtain the second upper bound in Eq.~(\ref{eq:gen_upperbound}), we
note that starting with a jump code $C=(N,K,d)_w$, applying $\sigma_x$
to all qubits yields a jump code $\overline{C}=(N,K,d)_{N-w}$. Note that,
$\sigma_x$ interchanges ground and excited state, hence the code
$\overline{C}$ lies in the decoherence free subspace with $N-w$
excitations. Also, the linear space spanned by the operators ${\cal
J}_E^\dagger {\cal J}_E$ for all subsets $E$ with no more than $d$
elements is invariant under conjugation by $\sigma_x$ on all
qubits. This holds as $\sigma_x L_\alpha^\dagger L_\alpha \sigma_x=
|0\rangle_{\alpha\alpha}\langle
0|=\openone-|1\rangle_{\alpha\alpha}\langle 1|$. Thus, for the code
$\overline{C}$ we obtain the bound $K\le{N-d \choose N-w-d}={N-d \choose
w}$.
If there is no restriction on the number $w$ of excited states,
choosing $w=\lfloor N/2\rfloor$ maximizes the upper bound of
Eq. (\ref{eq:gen_upperbound}), i.e.
\begin{equation}\label{maxupperbound}
K\le {N-d \choose \lfloor N/2\rfloor-d}.
\end{equation}
As mentioned above, the upper bound of Eq. (\ref{maxupperbound}) is
achieved for $d=1$ and for an even number $N$ of qubits. 
A table of lower bounds (obtained by constructions from $d$-${\rm
SEED}$s) and upper bounds for small values of $N$ and $d$ is provided
in \cite{DCC}.

\section{Stability properties of the one detected jump-error correcting quantum codes}\label{NonIdeal}
The detected jump-error correcting quantum codes constructed in the
previous sections can  stabilize quantum
algorithms provided three conditions are fulfilled.  First, the decay
rates of the qubits are equal.  Second, the time and position of each
quantum jump is detected with hundred percent efficiency. Third, the
appropriate unitary recovery operations are applied perfectly and
instantaneously immediately following the detection of an error.
Experimentally these requirements can only be approximated.
Therefore, the natural question arises how non-ideal conditions
influence the robustness of the embedded error-correcting quantum
codes.

In this section several types of imperfections affecting the ideal
performance of the one detected jump-error correcting embedded quantum
codes of Sec.~\ref{CodeDesign} are studied numerically. For this
purpose we investigate the stabilization of a quantum memory and of a
simple Hamiltonian dynamics against spontaneous decay processes. The
effective two-level Rabi Hamiltonian considered, i.e.
\begin{equation}
H = i \hbar \Omega~(|x_0 \rangle \langle v | - | v \rangle \langle x_0|) ,
\label{GrovHam}
\end{equation}
can be viewed as modeling the quantum dynamics of the ideal Grover
search algorithm \cite{Grover} in the limit of a large number of
qubits.  Thereby, the Rabi frequency $\Omega$ can be related to the
characteristic time $\tau_{op}$ required for performing an
oracle-operation and to the number of qubits $N$ according to $\Omega
= 2/(\tau_{op}\sqrt{N})$ \cite{Mussinger}.  Here, $|v\rangle$ denotes
an equally weighted superposition of all (orthonormal)
code words which may be used
as an initial state in Grover's quantum search algorithm.  The final
state we are searching for is denoted $|x_0\rangle$.  For this choice
of $|v\rangle$ the states $|v\rangle$ and $|x_0\rangle$ are not
orthogonal because $\langle v|x_0\rangle = 1/\sqrt{N}$. However, if
the number of qubits $N$ becomes large their overlap tends to zero.
According to the Hamiltonian Eq.~(\ref{GrovHam}) and consistent with
Grover's quantum algorithm \cite{Grover}, after the interaction time
$\tau = \pi /(2 \Omega)$, the initial state
$|v\rangle$ is transformed to the final state $|x_0\rangle$.

\subsection{Imperfect detection of error positions}

\begin{figure}[hbt]
\centerline{\psfig{figure=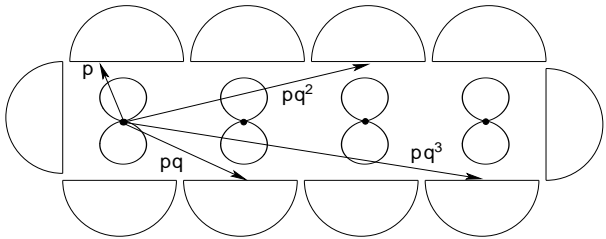,width=80mm}}
\caption{ Model for a 4-qubit register where jumps are detected with a given
error parameter $q$.}
\label{Fehldetekt}
\end{figure}

Let us first of all consider situation where the jump position can be
detected with a given non-zero error rate only. Such an imperfection might
occur if, for example, a photon emitted by a particular trapped ion 
is detected by the photodetector associated with a different ion (see
with Fig.~\ref{Fehldetekt}).

The probability to detect the emitted photon at the correct position
is denoted by $p$. The probability that an emitted photon is detected
falsely by the next nearest neighbor is given by $q \times p$.
Analogously, the probability of detecting the photon by the $n$-th
nearest neighbor is $q^n \times p$ with the normalization
condition $p \times ( 1+ q + q^2 + \ldots) \equiv 1 $.

\begin{figure}[hbt]
\centerline{\psfig{figure=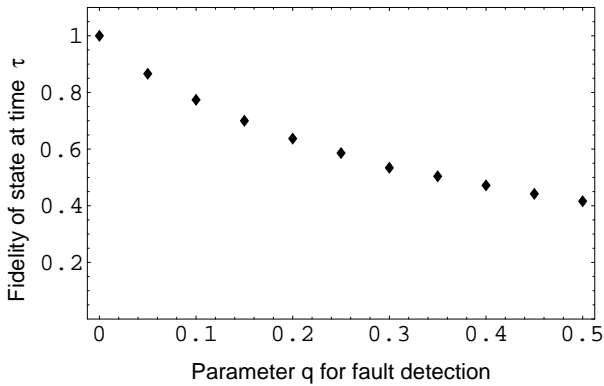,width=80mm}}
\caption{ Fidelity $ | \langle \psi ( \tau ) | \psi(0) \rangle |^2$ of a quantum memory after time $\tau = \pi/(2 \kappa)$ for an error parameter $q$.
\label{DetPosit}}
\end{figure}

The influence of this type of imperfection on a quantum memory,
i.e. $H\equiv 0$ in Eq.(\ref{master equation}), is depicted in Fig.~\ref{DetPosit}. A state $|\psi
(0) \rangle$ of the jump code $C=(4,3,1)_2$ is propagated according to
Eq.~(\ref{unravel}) with a Monte-Carlo simulation of the
quantum-trajectories \cite{Zoller}.  If a jump is detected, the appropriate recovery
operation is applied. In the case of a correct detection,
the quantum state of the memory is recovered perfectly.
In the case of a false detection, the quantum state of the memory leaves
the code space.  Therefore, in this simulation the full Hilbert space
of four physical qubits has to be taken into account.
As a measure of fidelity the squared absolute value of the overlap
 between
the state $|\psi (\tau)\rangle$ after 
a time $\tau = \pi / (2 \kappa) $ and the initial state $|\psi (0)
\rangle $ of the memory  is plotted as a function of the error parameter $q$.

\subsection{Unequal spontaneous decay rates}

For $\kappa_{\alpha}\equiv \kappa$ the orthonormal basis states ${|{\bm c}_i\rangle}$ of the code $ C \subset D$
are degenerate eigenvectors of the operator
$\sum_{\alpha}L^{\dagger}_{\alpha}L_{\alpha}$ appearing in the
effective Hamiltonian of Eq.~(\ref{effective Hamiltonian}). This property
ensures that these states form a passive code for the effective time
evolution between successive quantum jumps. The existence of such
degenerate eigenstates of the operator
$\sum_{\alpha}L^{\dagger}_{\alpha}L_{\alpha}$ relies on the assumption
that the decay rates of all qubits are equal.  Although this physical
situation can be realized in a laboratory, it is of interest to
investigate what happens if this condition of equal decay rates is
violated.  In this latter case our code does not correct errors
between successive quantum jumps passively.

For this purpose let us consider the Rabi-hamiltonian  of Eq.~(\ref{GrovHam})
which describes
the ideal quantum dynamics. In addition, we assume that
the decay rates of the physical qubits
are selected randomly according to a Gaussian distribution
whose mean value $\overline\kappa$ is
equal to the characteristic Rabi frequency $\Omega$.

\begin{figure}[hbt]
\centerline{\psfig{figure=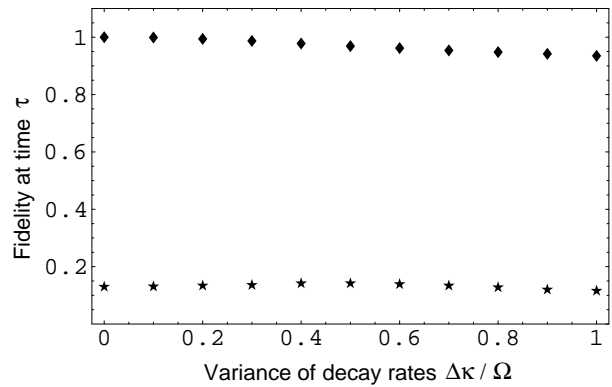,width=80mm}}
\caption{Dependence of the fidelity of an effective Rabi oscillation
at time $\tau = \pi / (2 \Omega) $ on the variance $\Delta \kappa$ of
the decay rates: The mean value of the decay rates is
$\overline{\kappa} = \Omega $, the variances are chosen from Gaussian
distributions with $\Delta \kappa = 0.1 \ldots 1.0 \times \Omega$. The
diamonds show the simulation using the one detected jump-error
correcting quantum code $(6,10,1)_3$, the stars show the corresponding results
for six qubits without encoding.}
\label{DeltaKappa}
\end{figure}

To study the resulting time evolution we choose the $(6,10,1)_3$ code
based on complementary pairings (see Sec.~\ref{CodeDesign}).
In
Fig.~\ref{DeltaKappa}
the fidelity of the quantum state $\rho(\tau)$ is depicted as a function of the variance $\Delta\kappa$ of the Gaussian distribution.
The fidelity is defined as the overlap 
$\langle x_0 | \rho (\tau) | x_0 \rangle $ between the (mixed) system state $\rho (\tau)$ at time $\tau = \pi /
(2 \Omega)$ and the desired state $ |x_0\rangle \langle x_0|$ which would result from the ideal dynamics
at this particular interaction time.
In this numerical simulation the master equation
~(\ref{master equation}) was integrated up to time $\tau = \pi / (2 \Omega)$,
whereas each jump operator $L_{\alpha}$ was replaced by a sequence
consisting of $L_{\alpha}$ and an immediately applied unitary recovery
operation $U_{\alpha}$ (see Eq.(\ref{recovery})). In this simulation
it was assumed that the recovery
operations are performed perfectly.
It is apparent that the code stabilizes the quantum dynamics
successfully, despite the fact that the $(6,10,1)_3$ code is not a
perfect one detected jump-error correcting quantum code for this
situation. 

%

\subsection{Time delay between error detection and recovery operation}

\begin{figure}[hbt]
\centerline{\psfig{figure=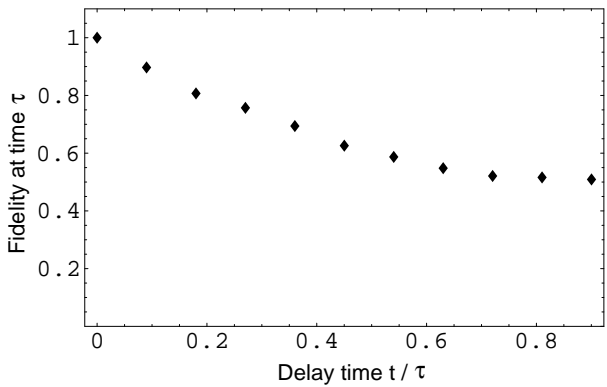,width=80mm}}
\caption{Fidelity $|\langle \psi (\tau) | x_0\rangle |^2$
for an effective Rabi oscillation 
at time $\tau = \pi / (2 \Omega) $  with  a delay time $t$ between
detection and recovery:
The decay rate is $\kappa=0.5 \; \Omega$.}
\label{DetDelay}
\end{figure}

Immediately after the detection of a spontaneous emission event the
qubits are described by a quantum state belonging to a subspace
involving one excited qubit less than the original code space. This
subspace also constitutes a passive error-correcting code. Therefore,
a time delay between the detection and the application of a recovery
operation does not lead to an additional error caused by the effective
time evolution $H_{\rm eff}$ provided the ideal quantum dynamics
characterized by $H$ is not affected. Nevertheless, this time delay
must be short in comparison with the mean time between two successive
spontaneous emission events. Otherwise, a second spontaneous emission
may map the state of the system onto another subspace, from which a
recovery is no longer possible.  Fig.~\ref{DetDelay} demonstrates
that, as long as the delay between detection and correction is not too
large compared with the mean decay time, error correction is still
possible.

\subsection{Dead times of the photodetectors}

\begin{figure}[thb]
\bigskip
\centerline{\psfig{figure=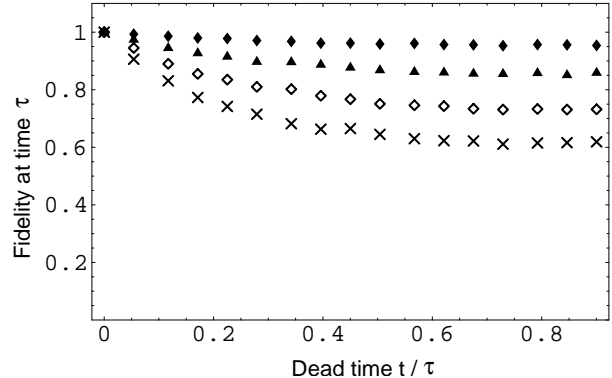,width=80mm}}
\caption{ Dependence of the fidelity $|\langle \psi (\tau) | x_0\rangle |^2$
at time $\tau = \pi / (2 \Omega)$ on the dead time $t$ of a
photodetector.  The ideal dynamics are governed by an effective Rabi
oscillation with Rabi frequency $\Omega$. For the encoding the one
detected jump-error correcting quantum code $(6,10,1)_3$ is used.  The
fidelity is depicted for four different values of $\overline{\kappa}$:
$\overline{\kappa} = 0.25 \;\Omega$ (black diamonds),
$\overline{\kappa} = 0.5 \; \Omega$ (triangles), $\overline{\kappa} =
0.75 \; \Omega$ (white diamonds), $\overline{\kappa} = 1.0 \; \Omega$
(crosses).}
\label{Totzeit}
\end{figure}

Another important condition for correct implementation of a detected
jump-error correcting quantum code is the ability to observe the
environment of each qubit {\em continuously}.  However, immediately
after the detection of a spontaneous emission event, typically the
detector is not able to respond to another photon. During the latent
response time of a photodetector a second spontaneous emission event
can take place which may destroy quantum coherence.  In
Fig.~\ref{Totzeit} the dependence of the fidelity $|\langle \psi
(\tau) | x_0\rangle |^2$ on the response time of a photodetector is
depicted for various decay rates $\kappa$.

It is apparent from Fig.~\ref{Totzeit} that the detected jump-error
correcting quantum code can stabilize an algorithm as long as the
response time of the photodetectors is small in comparison with the
average time between successive spontaneous emission events.

\section{Conclusions}
We have studied quantum error-correcting codes that exploit additional
information about the locations of the errors. This information is
obtained by continuously monitoring the system. Errors caused by the
resulting non-unitary dynamics are corrected passively by embedding
the error-correcting code in a decoherence free subspace.  To
construct such codes we have established connections to design theory.
The numerical simulations presented demonstrate that the jump codes
discussed can stabilize quantum systems even in cases of imperfect
detections and recovery operation.

\section{Acknowledgments}

This work is supported by the DFG (SPP
`Quanteninformationsverarbeitung') and by IST-1999-13021 and IST-2001-38869 of the
European Commission.

\end{document}